\documentclass[12pt]{article}
\usepackage{epsfig}
\newcommand{\secn}[1]{Section~\ref{#1}}
\newcommand{\bra}[1]{\langle{#1}|}
\newcommand{\ket}[1]{|{#1}\rangle}

\newcommand{\eq}[1]{Eq.~(\ref{#1})}

\def\ba{\begin{array}}
\def\ea{\end{array}}
\newcommand{\sect}[1]{\setcounter{equation}{0}\section{#1}}
\renewcommand{\theequation}{\thesection.\arabic{equation}}
\newcommand{\be}{\begin{equation}}
\newcommand{\ee}{\end{equation}}
\newcommand{\bea}{\begin{eqnarray}}
\newcommand{\eea}{\end{eqnarray}}

\def\dg{\dagger} 
\renewcommand{\hat}{\widehat}
\renewcommand{\tilde}{\widetilde}
\renewcommand{\a}{\alpha}
\renewcommand{\b}{\beta}
\def\g{\gamma} \def\G{\Gamma}
\renewcommand{\d}{\delta}
\newcommand{\e}{\eta} \def\p{\pi}
\def\m{\mu} \def\n{\nu} \def\s{\sigma}
\def\eps{\epsilon} \def\r{\rho} \def\f{\phi}
\def\l{\lambda} \def\z{\zeta} \def\t{\theta}
\def\L{\Lambda} \def\D{\Delta}
\def\cN{{\cal N}} \def\cV{{\cal V}}
\def\cA{{\cal A}} \def\cC{{\cal C}}
\def\cB{{\cal B}} 
\def\cF{{\cal F}} 
 
\def\cP{{\cal P}} 
\newcommand{\shalf}{\frac{1}{2}}

\newcommand{\NP}[1]{Nucl.\ Phys.\ {\bf #1}}
\newcommand{\PL}[1]{Phys.\ Lett.\ {\bf #1}}

\newcommand{\PRL}[1]{Phys.\ Rev.\ Lett.\ {\bf #1}}
\newcommand{\MPL}[1]{Mod.\ Phys.\ Lett.\ {\bf #1}}

\renewcommand{\thefootnote}{\fnsymbol{footnote}}
\def\one{{\hbox{ 1\kern-.8mm l}}}
\def\NS{{\rm NS}}
\def\R{{\rm R}}
\def\ii{{\rm i}}
\def\ee{{\rm e}}

\newlength{\bredde}
\def\slash#1{\settowidth{\bredde}{$#1$}\ifmmode\,\raisebox{.15ex}{/}
\hspace*{-\bredde} #1\else$\,\raisebox{.15ex}{/}\hspace*{-\bredde} #1$\fi}
\begin{document}
\begin{titlepage}
\rightline{DFTT 73/99}
%\rightline{hep-th/9903123}
\vskip 1.2cm
%\vskip 0.8cm
\centerline{\Large \bf Gauge and gravitational interactions}
\centerline{\Large \bf of non-BPS D-particles
\footnote{Work partially supported by the European Commission
TMR programme ERBFMRX-CT96-0045, by MURST
and by the Programme Emergence de la r\'egion Rh\^one-Alpes (France).
\\
E-mail:~~{\tt gallot,lerda,strigazzi@to.infn.it}}}
\vskip 1.5cm
\centerline{\bf L. Gallot $^a$,
A. Lerda $^{b,a}$ and P. Strigazzi $^a$}
\vskip .8cm
\centerline{\sl $^a$ Dipartimento di 
Fisica Teorica, Universit\`a di Torino}
\centerline{\sl and I.N.F.N., Sezione di Torino, 
Via P. Giuria 1, I-10125 Torino, Italy}
\vskip .4cm
\centerline{\sl $^b$ Dipartimento di Scienze e Tecnologie Avanzate}
\centerline{\sl Universit\`a del Piemonte Orientale, I-15100 
Alessandria, Italy}
\vskip 1.5cm
\begin{abstract}
\noindent
We study the gauge and gravitational interactions of the
stable non-BPS D-particles of the type I string theory. The
gravitational interactions are obtained using the boundary state
formalism while the $SO(32)$ gauge interactions are determined by 
evaluating disk diagrams with suitable insertions of boundary 
changing (or twist) operators.
In particular the gauge coupling of a D-particle is obtained from a
disk with two boundary components produced by the insertion of two
twist operators. We 
also compare our results with the amplitudes among the non-BPS 
states of the heterotic string
which are dual to the D-particles. After taking into account the known 
duality and renormalization effects, we find perfect agreement,
thus confirming at a non-BPS level the expectations based
on the heterotic/type I duality.
\end{abstract} 
\end{titlepage}
\newpage
\renewcommand{\thefootnote}{\arabic{footnote}}
\setcounter{footnote}{0}
\setcounter{page}{1}
\sect{Introduction}
\label{intro}
\vskip 0.5cm

The spectrum of supersymmetric string theories usually contains a special class
of states known as BPS states, which are characterized by the property
that their mass is completely determined by their charge under
some gauge field. They form short (or ultra short) supersymmetric multiplets
and, because of this fact, are 
stable and protected from quantum radiative corrections. A well-known
example of such BPS states is provided by the supersymmetric D$p$-branes
of the type II theories (with $p$ even in type IIA and $p$ odd in type IIB)
\cite{tasi}. However, supersymmetric string theories quite often contain
states that are stable without being BPS. These
are in general the lighest states which carry some conserved quantum numbers.
For them there is no particular relation between their mass and their charge;
they form long multiplets of the supersymmetry algebra and receive 
quantum radiative corrections.
However, being the lightest states with a given set of conserved quantum
numbers, they are stable since they cannot decay into anything else. 
Usually, it is not difficult to find such non-BPS states with the
standard string perturbative methods and analyze their properties at weak 
coupling; but, since they cannot decay, they
should be present
also in the strong coupling regime, or equivalently they
should appear as non-perturbative (D-brane type) configurations in the 
weakly coupled dual theory. To verify the existence of these non-BPS states 
is therefore a very strong test on the duality relations between 
two string theories which does not rely on supersymmetry arguments.
The study of the stable non-BPS D-branes in string theory, pioneered
by A. Sen in a remarkable series of papers \cite{sen1,sen3,sen4,sen5,sen6},
has attracted a lot of interest during the last year (for reviews see 
Refs.~\cite{sen7,rodolfo,SCHWARZ}) also for several other reasons; among
them we recall the fact the non-BPS D-branes might be useful for analyzing
the non-perturbative properties of the non-supersymmetric field theories
that live on their world-volumes, or the fact that they may lead to novel
types of relations among string theories \cite{sen7}.

One of the most notable examples of stable non-BPS configurations is
provided by the perturbative states at the first excited level 
of the $SO(32)$ heterotic string \cite{GHMR} which carry the spinor 
representation of the gauge group
and whose mass is given by
\begin{equation}
M_{\rm h} =  \frac{2}{\sqrt{\alpha'}} =  \frac{g_{\rm YM}}{\kappa_{10}}
~~.
\label{massh0}
\end{equation} 
In the last equality we have introduced the low-energy gauge and
gravitational couplings ${g_{\rm YM}}$ and ${\kappa_{10}}$ of the
heterotic string following the
conventions of Ref.~\cite{pol} which are also reviewed in Appendix B.
Being at the first massive level, these states are non-BPS, but being
the lightest ones carrying the spinor representation of $SO(32)$,
they are stable and should be present
also when one increases the heterotic string coupling constant $g_{\rm h}$.
In this process however, the mass $M_{\rm h}$ gets renormalized 
since there are no
constraints on it coming from supersymmetry. Thus we can write
\begin{equation}
M_{\rm h} =  \frac{g_{\rm YM}}{\kappa_{10}}~f
~~,
\label{massh}
\end{equation} 
where the renormalization function $f$ can in principle be computed
perturbatively in the heterotic string and is such that $f \sim 1$
for $g_{\rm h} \to 0$.  

If the heterotic/type I duality \cite{polwit} is correct, also in the type I
theory
there should exist 
stable non-BPS configurations that are
spinors of $SO(32)$. Such states do indeed exist and were 
identified by A. Sen as the non-BPS D-particles of type I \cite{sen4,sen5};
then, an explicit boundary state description for them was provided 
in Ref.~\cite{gallot}~\footnote{For the description 
of other non-BPS D-branes
using the boundary state formalism see 
Refs.~\cite{Gab1,Gab2,Gab3,Gab4,paris}.}. 
The mass of these D-particles turns out to be
\begin{equation}
M_{\tilde 0} =  \frac{1}{\sqrt{\alpha'}\,g_{\rm I}}
=\frac{g_{\rm YM}}{\kappa_{10}}~{2^{-3/4}\,{g_{\rm I}}^{-1/2}}
~~,
\label{massd0}
\end{equation} 
where $g_{\rm I}$, $g_{\rm YM}$ and $\kappa_{10}$ are, respectively, 
the string, the
gauge and the gravitational coupling constants of the type I theory 
in the conventions of Ref.~\cite{pol} (see also Appendix B).
Comparing Eqs. (\ref{massh}) and (\ref{massd0}), and remembering that
under the duality map the heterotic gauge and gravitational couplings turn
into the corresponding ones of type I, we can deduce that the renormalization
function $f$ must be such that 
$f \sim {2^{-3/4}\,{g_{\rm I}}^{-1/2}}$ for $g_{\rm I} \to 0$ in order
for the masses to agree on both sides. Clearly,
this result cannot be obtained using perturbative methods, but is a 
prediction of the heterotic/type I duality~\footnote{The T-dual 
heterotic/type I' correspondence has been analyzed at the non-BPS
level in Ref.~\cite{dasghupta}}. 

In this paper we elaborate further on these stable non-BPS particles
and study in detail their gravitational and gauge interactions.
On the heterotic side, these can be easily obtained
using standard perturbative techniques from correlation functions 
of vertex operators. In this way one can show, for example, that, at the lowest
order in the heterotic string coupling constant, the 
gravitational and gauge potential energies of two such particles at large
distance are given, respectively, by the Newton's law and the Coulomb's law
for massive and charged point-like objects in ten dimensions.
On the type I side, instead, the interactions of the 
non-BPS D-particles must be obtained using less standard methods
and have not been fully investigated 
so far; indeed,
only the general rules for computing string amplitudes with these 
D-particles have been given in the literature \cite{sen5,WITTEN}. 
It is the purpose of this
paper to fill this gap. 

In particular, we will concentrate on processes
involving massless string modes that are responsible for the
long range interactions among D-particles. 
To study the gravitational interactions we adopt the boundary state
formalism \cite{CLNY1,PCAI,cpb} and obtain the energy due to
the exchange of closed string states between two D-particles
by simply computing the diffusion amplitude between the two corresponding
boundary states in relative motion \cite{CANGEMI,06} (for recent reviews on 
the boundary state formalism and its applications see 
Ref.~\cite{paolo}). Then, by taking the
large distance limit to which only graviton and dilaton exchanges 
contribute, we find that the gravitational potential energy 
of two D-particles exactly agrees with the one of their heterotic duals, 
provided that the duality relations and the mass 
renormalization previously discussed are taken into account.
 
For the gauge interactions, instead, the situation is a bit more involved. 
In fact, we cannot use any more the boundary state formalism since this 
accounts only for the couplings of the D-particles with 
the closed strings that live in the bulk, but is completely 
blind to the other 
bulk sector of the type I theory consisting of open strings with Neumann 
boundary conditions in all directions to which the $SO(32)$ gauge fields
belong. On the other hand, the open strings attached to the D-particles have
Neumann boundary conditions only along the time direction. Therefore, to 
study the gauge interactions of our D-particles 
we should consider scattering amplitudes involving open strings with mixed
boundary conditions in an odd number of dimensions.
Calculations of open string amplitudes with mixed boundary conditions 
have already appeared in the analyses of systems of several
D-branes with different dimensionality (see for instance Ref.~\cite{Has}),
and require the use of twist operators to
produce mixed boundary conditions in certain directions.
These twist operators were used in the past to study
strings on orbifolds \cite{vafa}, and have been recently reconsidered 
from an abstract conformal field theory point of view \cite{reck}. 
Using such twist operators and 
applying the rules of Refs.~\cite{sen5,WITTEN}, we will describe how to 
compute scattering amplitudes involving non-BPS D-particles
and bulk open strings of type I. Special care is required in these
calculations because the twist operators that we use change the
boundary conditions in an {\it odd } number of directions. 
In particular, we will explicitly determine the gauge coupling of the 
D-particles by evaluating a
correlation function on a disk with two boundary components produced by 
the insertion of two twist operators. 
The result of this calculation is extremely simple, namely the 
non-BPS D-particles
couple minimally to the gauge field. 
Exploiting this fact, we then determine the gauge potential
energy of a pair of D-particles at large distance and see that 
after taking into account the duality map, this
exactly agrees with the corresponding energy computed in the heterotic
theory.

This paper is organized as follows: in \secn{gaugeamplitude}
we compute the gauge coupling of a
non-BPS D-particle of type I 
by evaluating a disk diagram with two twist
insertions, and then determine the gauge potential energy 
between two D-particles. 
In \secn{boundarystate} we use the boundary state formalism to
compute the gravitational contribution to the potential energy
of two (moving) D-particles. In \secn{heterotic} we study the gauge
and gravitational interactions of the non-BPS heterotic states that are
dual to the D-particles. In \secn{conclusions} we compare
the results for the non-BPS D-particles and for their dual heterotic
states, and discuss their relations. 
In Appendix A we show how to compute the gauge interactions between two
BPS D-strings of type I by extending the method of \secn{gaugeamplitude}
and verify the no-force condition. Finally, Appendices B and C contain
our conventions and a list of more technical formulas.    

\vskip 1.5cm
\sect{Type I D-particle interactions: the gauge amplitude}
\label{gaugeamplitude}
\vskip 0.5cm

As we mentioned in the introduction, an important check of the 
heterotic/type I duality has been the discovery by A. Sen 
\cite{sen3,sen4,sen5}
that the stable non-BPS heterotic states 
carrying the spinor representation of $SO(32)$
at the first massive level
are dual to the non-BPS D-particles of type I.
Specific rules for computing amplitudes involving such 
D-particles have been given by A. Sen \cite{sen5} and E. Witten \cite{WITTEN}
in two different ways which we briefly recall here. Sen's approach
heavily relies of the use of Chan-Paton factors to distinguish the
various kinds of open strings. The 0-0 
strings, whose end-points lie on the non-BPS D-particle, contain 
both states that are even and states that 
are odd under $(-1)^F$; the former carry a Chan-Paton factor $\!\one$, 
the latter a Chan-Paton factor $\sigma_1$. The 9-0 
strings stretching between one of the 32 D9 
branes of the type I background and a D-particle contain only
$(-1)^F$ even states but, due to the existence of an odd number of 
fermionic zero modes, their  vertex operators comprise the standard
GSO-even part as well as the corresponding GSO-odd part, weighted 
by Chan-Paton factors $\left( \begin{array}{c} 1\\0\end{array} \right)$ 
and $\left( \begin{array}{c} 0\\1\end{array} \right)$ respectively. 
Besides these factors,
the 9-0 strings also carry a Chan-Paton factor $\lambda^A$ 
($A=1,\dots,32$) labeling the fundamental representation of the
$SO(32)$ gauge group. Finally, the 9-9 strings are the usual open
strings of the type I theory which are GSO projected and carry
only the standard Chan-Paton factors of the gauge group.  
\par
The presence of the unusual
Chan-Paton factors $\!\one$, $\sigma_1$, 
$\left( \begin{array}{c} 1\\0\end{array} \right)$ or
 $\left( \begin{array}{c} 0\\1\end{array} \right)$ shows that
the states of the 0-0 and 9-0 sectors have a non trivial structure
which is really due to the presence of an {\it odd} number 
of fermionic zero modes. 
In order to remedy to this oddity, in Ref.~\cite{WITTEN} Witten has
proposed to introduce an extra one-dimensional fermion $\eta$  
on each boundary of the string world-sheet lying on a D-particle. In 
this way, in the 9-0 sector one recovers an even number of fermionic 
zero modes and can perform the usual GSO projection. Also in the 0-0
sector one performs a (generalized) GSO projection to obtain physical 
states, but since the extra fermion $\eta$ is odd under this GSO parity, 
one obtains two types of 0-0 states, similarly to what found by Sen.
\par
Let us now give some details on how to construct the massless
states in the various open string sectors using Witten's rules. 
We start with the 
NS sector of the 0-0 strings where at the massless 
level there are nine scalars $x^i$ ($i=1,\dots,9$) corresponding 
to the freedom of moving the D-particle in its nine tranverse 
directions. These
modes, which are present also on the
BPS D0 brane of the type IIA theory, correspond to vertex 
operators $\cV_{x^i}$ that do not depend on the boundary 
fermion $\eta$. 
In the $(-1)$
superghost picture, these vertex operators are simply
\begin{equation}
\cV_{x^i}^{(-1)} =  \psi^i~\ee^{-\phi}
%~~~,~~~V_{x^i}^{(0)} = \partial X^i 
~~.
\label{ns0}
\end{equation} 
Notice that there is no factor of $\ee^{\ii k\cdot X}$ in 
(\ref{ns0}) because massless states of $0$-$p$ strings have no 
momentum. Let us now consider the R sector of the 0-0 strings. Here 
both the ten world-sheet fermions $\psi^{\mu}$
and the boundary fermion $\eta$ possess zero modes
so that the massless R states 
form a GSO-even spinor $\xi^\alpha$ of $SO(1,10)$. Note that in this case
the GSO projection is simply the ten-dimensional chirality 
projection which is natural when one
extends $SO(1,9)$ to $SO(1,10)$ by adding $\eta$.
Thus, in the $(-1/2)$
superghost picture the vertex operator for the massless R states reads 
\begin{equation}
\cV_{\xi^\alpha}^{(-1/2)} = {1+\eta \over 2}\,S^{\alpha}~\ee^{-\phi/2}
\label{r0}
\end{equation} 
where $S^{\alpha}$ is the spin field of conformal dimension
$10/16$ associated to the ten  
world-sheet fermions. Upon quantization, the $16$ massless
fermionic modes described by (\ref{r0}) account 
for the $2^{16/2}=256$ degeneracy of the non-BPS D-particle. 

We now turn to the 9-0 strings which are more relevant for our purposes.
Since the NS sector does not contain massless states, we just consider
the R sector. In this case, the only world sheet fermion to have 
a zero mode is $\psi^0$ so that the ground state 
is a GSO-even (chiral) spinor of the algebra $SO(1,1)$ generated 
by $\psi^0_0$ and $\eta$. Hence, the vertex operator
describing the massless modes of the 9-0 sector should contain
\begin{itemize}
\item a spin field $S$ associated to the fermion $\psi^0$, of conformal 
dimension 1/16;
\item a boundary changing operator $\Delta$ for the nine space directions 
transverse to the D-particle, of conformal dimension 9/16;
\item a GSO (or chirality) projector for the Clifford algebra 
of $SO(1,1)$ ${1 + \psi_0^0 \eta\over 2}$, 
of conformal dimension zero;
\item a superghost contribution in the $-1/2$ picture $\ee^{-\phi/2}$, 
of conformal dimension 3/8;
\item a gauge Chan-Paton factor $\lambda^A$ to specify 
which of the
32 D9 branes one is considering. 
\end{itemize}
Thus, we have
\begin{equation}
\cV_{90}^{(-1/2)} = \lambda^A~ {1 + \psi_0^0\,\eta \over 2} ~S~ \Delta~ 
\ee^{-\phi/2}~~.
\label{r90}
\end{equation}
It is easy to check that the operator (\ref{r90}) has indeed conformal 
dimension 1 as it should be for a physical vertex operator. Notice that the
GSO projection in (\ref{r90}) keeps only one fermionic degree of freedom
for each value of the index $A$ of the fundamental representation 
of $SO(32)$. Upon quantization, 
the states described by 
$\cV_{90}$ form a spinorial representation of $SO(32)$, and 
hence
we can conclude that the marginal operator (\ref{r90}) accounts for
the $SO(32)$ degeneracy of the non-BPS D-particle. Since in type I the 
strings are unoriented, we should consider also the 0-9 sector. This is
merely related to the 9-0 sector through the action of the world-sheet
parity $\Omega$. Recalling that $\Omega$ simply acts by transposition on the
Chan-Paton factors without changing the physical content of the
vertex operators, we have 
\begin{equation}
\cV_{09}^{(-1/2)}= \Omega~\cV_{90}^{(-1/2)}= 
{\lambda^t}^{\,A}~ {1 + \psi_0^0\,\eta \over 2} ~S~ \Delta~ 
\ee^{-\phi/2}~~.
\label{r09}
\end{equation}
Notice in particular that the $SO(1,1)$ GSO projection is the same
in both vertices (\ref{r90}) and (\ref{r09}).

Finally, there are the 9-9 strings which, as we mentioned above, 
are the usual open strings of the type I theory; 
in particular in the
NS sector at the massless level we find the $SO(32)$ gauge bosons which are
described by the following vertex operators in the 
$(-1)$ superghost picture
\begin{equation}
\cV_{\rm gauge}^{(-1)} = \Lambda^{AB}~ A_{\mu}\,\psi^{\mu}~
\ee^{\ii k\cdot X}~\ee^{-\phi}
~~,
\label{vgauge}
\end{equation}
where $\Lambda^{AB}$ are the generators of $SO(32)$ in the
fundamental representation (see Appendix B for our conventions) and
$A_\mu$ is the polarization vector. 

We now face the problem of finding the coupling between the 
non-BPS D-particle and the gauge field.
Since the latter belongs to the 9-9 massless sector, the diagram we have 
to compute corresponds to a disk with a part of its boundary on the
D-particle and a part on the D9-branes from which the gauge boson
is emitted. This is represented in the Figure 1.
\begin{figure}[ht]
\label{disk}
\caption{\small The disk diagram describing the gauge 
coupling of a type I D--particle}
\vskip 0.5cm
\centerline{\epsfig{figure=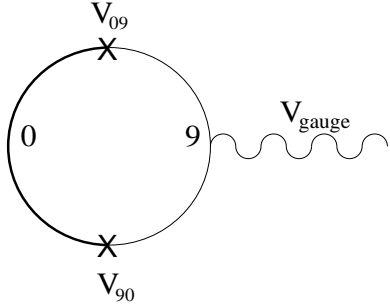,height=4cm,width=5.1cm,angle=0}}
\end{figure}
We thus have to insert two vertices containing the boundary changing
operator $\Delta$ which turns a boundary of type 0 into one of type 9
(or viceversa). The obvious choice is then to make insertions of the
vertices $\cV_{90}$ and $\cV_{09}$ given in (\ref{r90}) and (\ref{r09})
which  correspond to the $SO(32)$ degeneracy of the D-particle. Note
that, although no momentum is 
carried by these vertices, the emitted gauge boson may have non-zero 
space momentum. Indeed, the twist operators $\Delta$ are 
reservoirs of transverse momentum \cite{vafa} which allow emissions 
with non-zero momentum in the transverse directions; on the other 
hand, this is to be expected because the presence of a 
D-brane breaks the translational invariance in transverse space. 
Note also that, due to the insertion of the two vertices $\cV_{90}$ and 
$\cV_{09}$, the boundary component associated with 
the D-particle carries indices 
in the bi-fundamental representation of $SO(32)$. This is consistent with 
the fact that a D-particle should emit all, both massless and massive, 
perturbative open string states which group in the adjoint or in the 
symmetric representation of $SO(32)$. 

The gauge coupling of a (static) D-particle is then given by the 
expectation value of the gauge boson emission vertex (\ref{vgauge}) 
in the ``vacuum'' representing the D-particle. Thus, the diagram of Figure 1
corresponds to
\begin{equation}
{\cal F}^{\rm gauge} = 
\langle \Omega_{09} \vert\, \cV^{(-1)}_{\rm gauge} \vert \,\Omega_{90} \rangle
\label{vev}
\end{equation}
where 
\begin{equation}
\vert \Omega_{90} \rangle 
= \lim_{z\rightarrow 0} \cV^{(-1/2)}_{09}(z)\, \vert 0 \rangle 
~~~{\rm and}~~~\langle \Omega_{09} \vert 
= \lim_{z\rightarrow \infty} \bra{0}\,\cV^{(-1/2)}_{09}(z)~~. 
\end{equation}
Note that due to the presence of the twist operators in ${\cV}_{09}$ and 
${\cV}_{90}$, the expectation value of the gauge emission vertex
is not vanishing, as we will explicitly see in the following.

After including the normalization factor
${\cal C}_{\rm disk}$ appropriate of any disk amplitude, the 
normalization factors ${\cal N}_\R$ 
for the R vertices (\ref{r90}) and (\ref{r09}), and 
${\cal N}_\NS$ for the NS vertex (\ref{vgauge})~\footnote{We refer to 
Appendix B for the explicit expression of these normalization factors 
and to Refs.~\cite{pol,1loop} for their derivation.}, ${\cal F}^{\rm gauge}$
may be reexpressed as a 3-point function on the world sheet and reads
\begin{equation}
{\cal F}^{\rm gauge} = {\cal C}_{\rm disk}~{\cal N}^2_\R ~{\cal N}_\NS~
\int d\mu(z_i) ~\left
\langle 
c\,\cV^{(-1/2)}_{09}(z_1)~c\,\cV^{(-1)}_{\rm 
gauge}(z_2)~c\,\cV^{(-1/2)}_{90}(z_3) \right \rangle_{\eta} ~~.
\label{f3}
\end{equation}
where we have also added a ghost $c$ in each vertex operator. The 
notation $\langle~\rangle_{\eta}$ means that the correlator must be 
evaluated by including the action for the boundary fermion $\eta$
as explained in Ref.~\cite{WITTEN}. 

The correlation function in (\ref{f3}) 
may be decomposed into a longitudinal and a transverse piece. The latter
vanishes because
\begin{equation}
\Big\langle  \Delta(z_1)~\psi^i(z_2) ~\ee^{\ii k\cdot X(z_2)}
~\Delta (z_3)\Big\rangle =0
\label{transv}
\end{equation}
for $i=1,\ldots,9.$
Thus, there is no emission of gauge bosons with polarization $A_i$
along the transverse directions, as it should be for a minimally coupled
particle at rest.
We then consider the longitudinal part for which the basic correlators
are
\begin{eqnarray}
\Big\langle  c(z_1)\,c(z_2)\,c (z_3)\Big\rangle &=& 
z_{12}\,z_{13}\,z_{23}~~,\label{ghost}\\
\Big\langle \ee^{-\phi(z_1)/2}\,\ee^{-\phi(z_2)}\,\ee^{-\phi(z_3)/2}
\Big\rangle 
&=& z_{12}^{-1/2}\,z_{13}^{-1/4}\,z_{23}^{-1/2}~~,\label{superghost}\\
\Big\langle \Delta(z_1) \,\ee^{\ii k\cdot X(z_2)}\,
\Delta(z_3)\Big\rangle 
&=&z_{12}^{-\alpha' k^2}\,z_{13}^{\alpha' k^2-9/8}\,z_{23}^{-\alpha' 
k^2} ~~.\label{delta}
\end{eqnarray}
Notice that in (\ref{delta}) the transverse momentum $k^i$ of the 
emitted gauge boson is not subject to any constraint, as we have 
anticipated. The remaining correlator to be considered is 
\begin{equation}
\Big\langle \left({1 + \psi^0_0\,\eta \over 2} S(z_1)\right)  
~\psi^0(z_2)~
\left( {1 + \psi^0_0\,\eta \over 2} S(z_3) \right)\Big\rangle_{\eta} ~~.
\label{s1}
\end{equation}
This splits into four pieces, two of which vanish. Indeed, according to
Ref.~\cite{WITTEN} the only non-vanishing correlation functions are those
containing {\it one} factor of $\eta$. In particular one has
\begin{equation}
\Big \langle \eta\Big\rangle_{\eta} = \sqrt{2}~~~, ~~~  
\Big\langle 1 \Big\rangle_{\eta} = 0~~.
\label{eta} 
\end{equation}
Finally, we have
\begin{equation}
\Big\langle \psi^0_0 S(z_1) ~\psi^0(z_2)~ S(z_3) \Big\rangle  = 
\Big \langle  S (z_1) ~\psi^0(z_2) ~\psi^0_0 S(z_3)\Big\rangle =
z_{12}^{-1/2}\,z_{13}^{3/8}\,z_{23}^{-1/2}
~~.
\label{spin}
\end{equation}
Notice that a correlation function similar to (\ref{spin}) appears in 
the 2D Ising model. Indeed, the spin field $S$ may be identified with 
the order parameter $\sigma$ ({\it i.e.} the magnetization) while
the other spin field $\psi^0_0 S$ plays the role of the disorder 
parameter $\mu$ \cite{BPZ}. 

Inserting Eqs. (\ref{ghost})-(\ref{spin}) into (\ref{f3}) and exploiting 
the projective invariance to fix the position of the three punctures at
arbitrary values, we easily get
\begin{eqnarray}
{\cal F}^{\rm gauge} &=& -\,
{{\cal C}_{\rm disk}\,{\cal N}_\R^2\,{\cal N}_\NS \over
\sqrt{2}} ~{\rm Tr}({\lambda^t}^A\,\Lambda^{BC}\,\lambda^D)
~A_0~\int d\mu(z_i)~\left({z_{13}\over z_{12}z_{23}} 
\right)^{\alpha' k^2} 
\nonumber \\
&=& -\,{{\cal C}_{\rm disk}\,{\cal N}_\R^2\,{\cal N}_\NS \over
\sqrt{2}} ~{\rm Tr}({\lambda^t}^A\,\Lambda^{BC}\,\lambda^D)
~A_0~~.
\label{f31}
\end{eqnarray}
Then, using the explicit expressions of the normalization coefficients
and Chan-Paton factors given in Appendix B, we can rewrite 
${\cal F}^{\rm gauge}$ as follows 
\begin{equation}
{\cal F}^{\rm gauge} = -\,\ii ~{g_{\rm YM} \over \sqrt{2}} 
~\left(\delta^{AB}\,\delta^{CD}-\delta^{AC}\,\delta^{BD}\right)~A_0~~,
\label{f32}
\end{equation}
where $g_{\rm YM}$ is the gauge coupling constant of the type I theory.
 
Eq. (\ref{f32}) 
represents the amplitude for the emission of a gauge
boson with longitudinal polarization $\xi_0$ and 
color index $(BC)$ from a 0-boundary in the bi-fundamental of $SO(32)$.
The appearance of this representation is a direct consequence
of our construction in which the D-particle is represented
by the 0-component of a disk boundary produced by
the insertion of the vertex operators ${\cal V}_{90}$ and 
${\cal V}_{09}$.
On the other hand, the $SO(32)$ spinor degeneracy of the non-BPS
D-particle of type I arises from the (second) quantization of the
32 massless fermionic zero-modes of the 0-9 open strings, and thus
it is clear that such a degeneracy cannot be seen in our
operator formalism. This fact should not be surprising because
a completely analogous situation occurs in the familiar description
of D$p$-branes using boundary states. Indeed, a boundary state is a 
{\it single} state that correctly represents a D$p$-brane and its couplings
to the bulk closed strings, even if it
does not account for the degeneracy of the D$p$-brane under the
supersymmetry algebra. Similarly, in our case the 
0-component of the disk boundary produced by the insertion of 
${\cal V}_{90}$ and ${\cal V}_{09}$ correctly describes a D-particle and 
allows to obtain its coupling 
with the bulk 9-9 open strings, even if does not account for its
degeneracy under the gauge group. In fact, as we will see later 
and in the following sections, using this construction
we are able to obtain non trivial information about the gauge
interactions between two D-particles at large distance. Moreover, 
after taking into account the known duality relations, we will show
that the results obtained in this way exactly agree with those
in the heteroric theory, as required by the heterotic/type I duality, 
thus confirming the validity of our construction.

Using the result (\ref{f32}) we can now easily compute the gauge
potential energy ${V}_{\rm I}^{\rm gauge}$ due to the exchange of the $SO(32)$
gauge bosons between two D-particles. As indicated in Figure 2, 
\begin{figure}[ht]
\label{diskdisk}
\caption{\small The diagram describing the gauge amplitude
between two D-particles}
\vskip 0.5cm
\centerline{\epsfig{figure=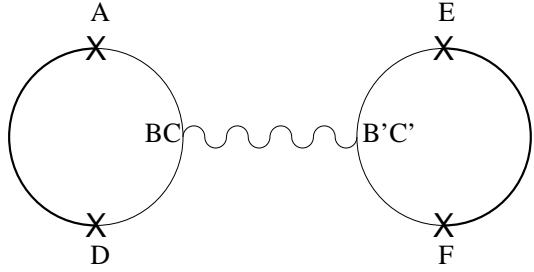,height=3.5cm,width=7cm,angle=0}}
\end{figure}
this can
be obtained simply by sewing
two emission amplitudes $\cF^{\rm gauge}$ with the gauge boson propagator 
\begin{equation}
\cP = {\e^{\a \b} \over q^2}~\left( \d^{BB'}\d^{CC'}- \d^{BC'}\d^{CB'} 
\right) ~~,
\label{gaugebos}
\end{equation} 
yielding 
\begin{equation}
{V}_{\rm I}^{\rm gauge} = 
 - \,\frac{g_{\rm YM}^2}{2}~\left(\delta^{AE}\,\delta^{FD}
-\delta^{AF}\,\delta^{DE}\right)~\frac{1}{q^2}~~.
\label{vgauge00}
\end{equation}
Performing a Fourier transform, we get 
the following (static) gauge potential in configuration space
\begin{equation}
{V}_{\rm I}^{\rm gauge}(r) =
-\, \frac{g_{\rm YM}^2}{2}~\left(\delta^{AE}\,\delta^{FD}
-\delta^{AF}\,\delta^{DE}\right)~\frac{1}{7\,\Omega_8\,r^7}~~,
\label{vgauge0}
\end{equation}
where $\Omega_q=2\pi^{(q+1)/2}/\Gamma((q+1)/2)$ is
the area of a unit $q$-dimensional sphere. Eq.~(\ref{vgauge0}) clearly
represents a ``Coulomb-like'' potential energy 
for point particles at a distance $r$ in ten dimensions.

We conclude this section by mentioning that the same results
(\ref{f32}) and (\ref{vgauge0}) can be obtained also using the
rules given by A. Sen in Ref.~\cite{sen5} for computing amplitudes 
with non-BPS D-particles.

\vskip 1.5cm
\sect{Type I D-particle interactions: the gravitatio-
nal amplitude}
\label{boundarystate}
\vskip 0.5cm

The gravitational contribution to the scattering of two non-BPS D-particles 
of type I can be calculated,
at the leading order in the string coupling constant, from the 
diffusion amplitude between two corresponding boundary states. The 
boundary state description of the non-BPS D-particles has already been given
in Refs.~\cite{sen5,gallot} from which we recall the results
that are relevant in the forthcoming analysis. 
For details and conventions on boundary states, we refer the reader to 
Refs.~\cite{cpb,06,gallot}. 

In the closed string operator formalism, one describes
a D$p$-brane by means of a boundary state $\ket{Dp}$
\cite{CLNY1,PCAI}. This is
a closed string state which
inserts a boundary on the world-sheet, enforces on it the appropriate 
boundary conditions
and represents the source for the closed strings emitted
by the brane. 
As an example, the boundary state for a BPS D-particle of type IIA may 
formally be written as~\begin{footnote}{In order to avoid clutter, 
we shall denote  the NS-NS (resp. R-R) component of a boundary state 
with the simplified subscript NS (resp. R)}\end{footnote}
\begin{equation}
\ket{D0}_{\rm{IIA}}= \ket{D0}_{\NS}+\ket{D0}_{\R}~~,
\label{d02a}
\end{equation}
where the NS-NS and the R-R components are both
proportional to $T_0$ which is the tension 
of the D-particle in units of the gravitational coupling 
constant, namely
\begin{equation}
T_0= 8\,\pi^{7/2}\,\alpha'^{\,3/2}~~.
\label{t0}
\end{equation}
The presence of both the NS-NS and the R-R components implies that 
the spectrum of the open strings living 
on the D-particle is GSO-projected. The partition 
function of such open strings may be obtained by evaluating the 
cylinder/annulus amplitude in the closed string channel which is given by
\begin{equation}
{}_{\rm{IIA}}\bra{D0} P \ket{D0}_{\rm{II}A}~~,
\label{2a}
\end{equation}
and then performing a modular transformation. In Eq. (\ref{2a}),
$P$ denotes the closed string propagator
\bea
P = {\a' \over 2}\,{1 \over L_0 + \tilde{L}_0 -a-\tilde{a}}~~.
\label{propagator}
\eea
where $a$ ($\tilde{a}$) is the left (right) intercept 
($a_{\NS} = 1/2$, $a_{\R}=0$).

The boundary state for the non-BPS D-particle of the IIB theory 
\cite{sen3,gallot} 
has instead only a component along the NS-NS sector and a tension 
$T_{\tilde{0}}$ greater by a factor of $\sqrt{2}$ than 
$T_{0}$. Thus, we can write
\begin{equation}
\ket{D\tilde{0}}_{\rm{IIB}}= \sqrt{2}\ket{D0}_{\NS}~~.
\end{equation}
As a consequence, there is no GSO-projection in the spectrum of the 
open strings lying on the non-BPS D-particle and the presence of 
a tachyon in the NS sector renders it unstable. 
However, if we consider the type I theory \cite{ps},
the tachyon is removed by the projection onto states invariant under the
world-sheet parity $\Omega$ \cite{sen3,gallot}. 
In the boundary state formalism,
the $\Omega$ projection is implemented by adding 
the so-called crosscap state $\ket{C}$ \cite{PCAI}, which corresponds
to inserting on the closed string world-sheet a boundary 
with opposite points identified. 
The negative $(-32)$ charge
for the non-propagating R-R 10-form that the crosscap 
generates in the background, must
be canceled 
by the introduction of 32 D9-branes. Hence, the type I theory 
possesses a background ``boundary state'' given by
\cite{PCAI} 
\begin{equation}
{1 \over \sqrt{2}} \Big(\ket{C}+32\ket{D9} \Big)
\label{background}
\end{equation}
where the factor of $1/\sqrt{2}$ has been introduced to obtain the right 
normalization of the various spectra. 
Then, the partition function for unoriented open 9-9 strings, given 
by the sum of the annulus and the M\"obius strip contributions, is
\begin{equation}
{1 \over 2} \left(2^{10} \bra{D9}P\ket{D9} +
2^5 \bra{D9}P\ket{C} + 2^5\bra{C}P\ket{D9}\right)~~,
\end{equation}
while the contribution of the Klein bottle
\begin{equation}
{1 \over 2}  \bra{C}P\ket{C}
\end{equation}
added to the torus contribution gives 
the partition function for unoriented closed strings.

The boundary state of the non BPS D-particle of type I reads
\begin{equation}
\ket{D\tilde{0}}_{I}= 
{1 \over \sqrt{2}}\left(  \sqrt{2} \ket{D0}_{\NS} \right) =\ket{D0}_{\NS}
\end{equation}
where we have added the same factor of $1/\sqrt{2}$ for 
consistency with (\ref{background}). 
The mass of the D-particle is then given by
\bea
M_{\tilde{0}} =\frac{1}{\sqrt{2}}\,{T_{\tilde{0}} \over \kappa_{10}} 
= {T_{0} \over \kappa_{10}}={1 \over \sqrt{\a'} \,g_I}
\label{tau0}
\eea
where $\kappa_{10}$ is the ten dimensional gravitational coupling 
constant of the type I theory (see Appendix B).
The partition function for open 0-0 strings living on the D-particle,
obtained by summing the contributions from the annulus and the M\"obius strip,
is
\begin{equation}
{}_{\NS}\bra{D0}P\ket{D0}_{\NS}+
{1 \over 2}\Big(\sqrt{2}\,\bra{C}P\ket{D0}_{\NS}+
 \sqrt{2}~{}_{\NS}\bra{D0}P\ket{C} \Big) ~~.
\end{equation}
In this theory, there are also 0-9 and 9-0 open strings with  one end on 
the D-particle and 
the other on one of the 32 D9 branes of the type I background.
The world-sheet parity $\Omega$ exchanges the two sectors 0-9 and 9-0 
so that we only retain
symmetric combinations corresponding to the partition function
\begin{equation}
{32\sqrt{2} \over 2}\Big( {}_{\NS}\bra{D0}P\ket{D9}+
\bra{D9}P\ket{D0}_{\NS}\Big)~~.
\end{equation}
The spectrum of open strings stretching between two {\it different} 
(distant) D-particles at rest, 
one labeled with a prime, has a partition function given by
\begin{equation}
{1\over 2} \left((\sqrt{2})^2 {}_{\NS}\bra{D0}P\ket{D0'}_{\NS}+
(\sqrt{2})^2{}_{\NS}\bra{D0'}P\ket{D0}_{\NS}\right)
\label{00'}
\end{equation} 
where the factor of one-half indicates that, compared to the IIB case, 
only the $\Omega$ symmetric combinations are retained. 
Notice that, at sufficiently small
distance, a tachyon develops in this open string spectrum signaling 
the instability of the configuration which decays into the vacuum \cite{sen5}.

Our aim is to study the diffusion of a moving D-particle 
with a velocity $v$ along
one space direction, say $X^1$, on another D-particle at rest at the origin. 
Such an interaction may be evaluated analyzing the spectrum of the open 
strings stretching between the two objects
with modified boundary conditions in the 0 and 1 directions. This can be
done generalizing the treatment for the BPS D-branes presented in
Ref.~\cite{Bachas}, but we find it simpler to use the  
method of the boosted boundary state \cite{CANGEMI,06}. Indeed, the 
interaction amplitude just reads
\begin{equation}
\cA_{\rm I}(v) = {}_{\NS}\bra{D0'} P\, \L \ket{ D0}_{\NS}+ 
{}_{\NS}\bra{D0} \L^{\dagger} \,P\ket{ D0'}_{\NS}
\label{scatteringI}
\end{equation}
where $\L$ is the boost operator
\begin{equation}
\L = \ee^{ \ii \p \n J^{01}}
\end{equation}
acting on the boundary state of a particle at rest. Here we 
have $v = \rm{th}(\pi \n)$
and $J^{\m \n}$ is the generator of the Lorentz transformations. 
Notice that the amplitude 
(\ref{scatteringI}) reduces to the static one (\ref{00'}) in the limit of 
vanishing velocity.
The boosted boundary state
\begin{footnote}{The signs $\pm$ correspond to the two possible 
implementations of boundary conditions for world-sheet fermions. In a 
physical (GSO projected) boundary state, 
only a suitable linear combination of them is retained.}
\end{footnote}
reads 
\bea
\L \ket{D0, \pm}_{\NS} &=& \frac{T_0}{2} \,{1\over \gamma} \,
\d^{(8)}(x) \,\d(x^0 \,v + x^1)\,
 \exp \left[ - \sum_{n=1}^{\infty} a_n^{\dg} \cdot S \cdot
\tilde{a}_n^{\dg} \right] \nonumber \\
&&\times\,
\exp \left[ \pm \,\ii \sum_{r=\shalf}^{\infty} \psi_r^{\dg} \cdot S \cdot
\tilde{\psi}_r^{\dg} \right] \prod_{\m = 0}^9 \ket{k^{\m} = 0}
\label{D0NSdyn}
\eea
where the boundary conditions are encoded in the matrix 
$S = (- \cV_{01} , -\one_8)$ with
\bea
\cV_{01} = \left(
 \ba{cc}
 \cosh (2\pi\n) & \sinh (2\pi\n) \\
 \sinh (2\pi\n) & \cosh (2\pi\n) \\
 \ea
\right) ~~.
\eea
Note that $\cosh(\pi\n)\equiv\gamma$ is the Lorentz factor. 

The interaction amplitude $\cA_{\rm I}(v)$ can be evaluated using
standard techniques \cite{CANGEMI,06} and explicitly reads
\begin{footnote}{See for instance Ref.~\cite{KIRITSIS} 
for definitions and conventions about the modular 
functions $f_k$ and $\theta_k$.}
\end{footnote}
\bea
\cA_{\rm I}(v) &=&  (8 \p^2 \a')^{-\frac{1}{2}}  \,
\int_{- \infty}^{\infty} d \tau
\int_0^{\infty} ds \,\,  s^{-\frac{9}{2}} \,\,
{\rm e}^{- { b^2+\tau^2 v^2\gamma^2 \over 2 \p \a' s}} \nonumber \\  
&&\,2 \,\sinh (\pi \n)\left[ {f_3^6(q) \,\t_3(-\ii\n|\ii s) - f_4^6(q) 
\,\t_4(-\ii\n|\ii s) \over f_1^6(q)\, \t_1(-\ii\n|\ii  s) } \right]
\label{scattering1}
\eea
in which $q=\ee^{-\p s}$, $\tau$ is the proper time of the moving 
particle and $b$ is the impact parameter.
We are now in a position to extract the long range interaction 
potential energy $V_{\rm I}^{\rm grav}$ due to gravitational
exchange between the two particles. To do so we have to perform 
the limit $s \rightarrow \infty$ in the 
integrand of \eq{scattering1}, then integrate on the variable $s$ and 
finally identify the potential energy according to
\begin{equation}
\cA_{\rm I}(v) = \int_{- \infty}^{\infty} d \tau \,\,V_{\rm I}^{\rm grav}
 ~~.
\end{equation}
In the non relativistic limit, we obtain the Newton's law with its first 
correction 
\begin{equation}
V_{\rm I}^{\rm grav}(r) =  
(2 \kappa_{10})^2 \,{M_{\tilde{0}}^2 \over 7\, \Omega_8\, r^7}
\left(1 + \shalf \,v^2\right) +o(v^2) \qquad v \rightarrow 0
\label{newton1}
\end{equation}
where we have introduced the radial coordinate $r^2 = b^2+v^2 \g^2 \tau^2$. 
Thus, in the non relativistic limit the boundary state calculation 
reproduces correctly the gravitational potential energy that we expect for a
pair of D-particles in relative motion.

\vskip 1.5cm
\sect{Interactions of the heterotic non-BPS states}
\label{heterotic}
\vskip 0.5cm

The non-BPS D-particles described in the previous sections account 
for the presence in the spectrum of the
type I theory of long super-multiplets of states carrying the spinorial 
representation of $SO(32)$. These non-perturbative states are dual to those 
appearing at the first massive level in the heterotic theory. 
Carrying the same quantum numbers, one naturally expects that these heterotic
states have the same kind of interactions as the D-particles of type 
I. In this section we will check this idea and investigate the 
gauge and gravitational interactions of the non-BPS heterotic states
using standard tools of perturbative string theory. In doing so, 
we will adopt the bosonized formulation of the heterotic string in which 
the gauge degrees of freedom are described by sixteen chiral bosons 
${\tilde X}^I$ ($I=1,\cdots,16$) appropriately compactified \cite{GHMR}.

The long super-multiplet of the stable heterotic states appears at the first 
massive level $\Big(M_{\rm h}^2 = 4/\a'\Big)$, and contains 
the following bosonic states 
\bea
 \psi_{-3/2}^{\mu} 
\ket{k} \otimes \ket{K^I}~~~&,& ~~~
\a_{-1}^{(\mu } \psi_{-1/2}^{ \nu)}
\ket{k} \otimes \ket{K^I} ~~,
\label{alpha}
\\
\a_{-1}^{[\mu } \psi_{-1/2}^{ \nu]} 
\ket{k} \otimes \ket{K^I}~~~ &,&~~~
\psi_{-1/2}^{\mu} \psi_{-1/2}^{\nu}\psi_{-1/2}^{\rho}
\ket{k} \otimes \ket{K^I}~~.
\label{gamma}
\eea
with $\mu,\nu,...=0,\ldots,9$. In these formulas $k$ denotes the
space-time momentum ($k^2=-M_{\rm h}^2$) while $K^I$ is the adimensional
momentum associated to the sixteen internal coordinates ${\tilde X}^I$. 
The states of Eq. (\ref{alpha}) describe massive degrees of freedom which
transform in the {\bf 44} representation of the Lorentz group, whereas those 
of Eq. (\ref{gamma}) transform in the {\bf 84}~\footnote{
The fermionic states that complete this long multiplet 
transform in the {\bf 128} representation of the Lorentz group.}. 
The level matching condition requires  
that $K^2=4$. This may be realized for example 
by taking $K^{I}$ to be of 
the form $(\pm {1 \over 2}, \pm {1 \over 2},, \cdots , \pm {1 \over 2})$ 
with an even number of $+$ signs, thus obtaining the spinorial 
representation of $SO(32)$ with positive chirality. 
The vertex operators for the 
states (\ref{alpha}) and (\ref{gamma}) will be denoted by $\cal V$ 
and can be found in Appendix C both in the 
$(-1)$ and in the $(0)$ superghost pictures.

We now study the interactions of these states with the massless
gauge bosons of $SO(32)$. In the bosonized formulation of the 
heterotic string we must distinguish between the states associated
to the $16$ Cartan generators that are given by 
\begin{equation}
A_\mu\,\psi_{-1/2}^{\mu} \ket{q} \otimes \tilde{\a}_{-1}^I \ket{Q=0}
\quad \mbox{with}~~q^2=0 \quad \mbox{and}~~I=1,\ldots,16~~,
\label{cartan}
\end{equation}
and those associated to the remaining $480$ generators which are
instead given by 
\begin{equation}
A_\mu\,\psi_{-1/2}^{\mu} \ket{q} \otimes \ket{Q}\quad \mbox{with}~~q^2=0
\label{charged}
\end{equation}
and the internal momentum $Q$ of the form 
$(0, \cdots , 0, \pm 1,0, \cdots , 0, \pm 1 , 0, \cdots , 0)$.
Also the vertex operators for the states (\ref{cartan}) and (\ref{charged}),
which we denote collectively by $\cV_{\rm gauge}$, can be found in Appendix
C in the $(-1)$ and $(0)$ superghost pictures.

The gauge coupling of the states (\ref{alpha}) and (\ref{gamma})
is obtained by simply computing the 3-point function on the sphere
among two vertex operators $\cV$ and one vertex operator
$\cV_{\rm gauge}$ (see Figure 3).
\begin{figure}[ht]
\label{three}
\caption{\small The 3-point function on the sphere}
\vskip 0.3cm
\centerline{\epsfig{figure=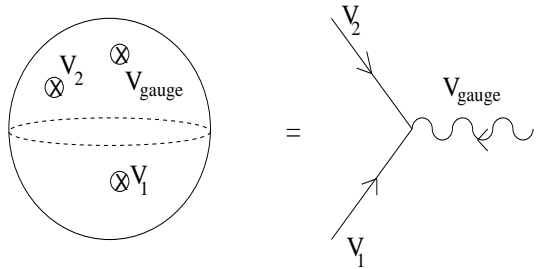,height=3.5cm,width=7cm,angle=0}}
\end{figure}
Including the normalization factor $\cC_0$ appropriate
of any tree-level closed string amplitude and a normalization
factor $\hat{\cN}$ for each vertex operator, we have
\begin{equation}
\Big( \cF_3^{\rm gauge} \Big)_{\a_1\a_2} =
\cC_0\, \hat{\cN}^3 \int d^2\m(z_i, \bar{z}_i) 
\,\Big\langle c{\bar c}\cV^{(-1)}_{1}(z_1,\bar{z}_1)\,
c{\bar c}\cV^{(-1)}_{2}(z_2,\bar{z}_2)\,
c{\bar c} \cV^{(0)}_{\rm gauge}(z_3,\bar{z}_3)
\Big\rangle 
\label{fgauge}
\end{equation}
where $\a_1$ and $\a_2$ label the spinor representation of $SO(32)$
carried by the non-BPS states and a ghost factor $c{\bar c}$ has been 
added in each puncture.
Actually, we are not interested in the complete expression of this
correlation function but only in the scalar part of it, namely
in the terms where the polarizations $\z_1$ and $\z_2$
of the two spinor states are contracted between 
themselves~\footnote{The polarization $\z_i$ can be either a vector, a
symmetric or antisymmetric two-index tensor or an antisymmetric three-index 
tensor depending on which particular states 
(\ref{alpha}) and (\ref{gamma}) are considered.}.
This is because we want to compare our results with those of the
non-BPS D-particles of type I obtained in the previous sections in
which the Lorentz group structure was not manifest. 

Using the explicit expression of the vertex operators reported
in Appendix C, we find that the terms of (\ref{fgauge}) proportional
to $\z_1\cdot\z_2$ are given by 
\begin{equation}
\Big( \cF_3^{\rm gauge} \Big)^I_{\a_1\a_2} =
 {\cC_0 \,\hat{\cN}^3\, \sqrt{2\a'} \over 4}\,
(\z_1 \cdot \z_2)\,A_{\mu} \,(k_1^{\mu}-k_2^{\mu})
\,  (K_{\a_1} - K_{\a_2})^I\,
\d_{K_{\a_1}+K_{\a_2},0}
\label{cartan1}
\end{equation}
when $\cV_{\rm gauge}$ corresponds to the gauge bosons associated to the 
Cartan generators, and
by 
\begin{equation}
\Big( \cF_3^{\rm gauge} \Big)^Q_{\a_1\a_2}= 
{\cC_0\, \hat{\cN}^3\, \sqrt{2\a'} \over 2}\,
(\z_1 \cdot \z_2)\,A_{\mu}\, (k_1^{\mu}-k_2^{\mu}) 
\,C_Q(K_{\a_1}) \,\d_{K_{\a_1}+K_{\a_2}+Q,0}
\label{fgauge1}
\end{equation}
when $\cV_{\rm gauge}$ corresponds to the gauge bosons associated to the 
remaining generators
(here $K_{\a}$ denotes the lattice vector corresponding to the spinorial 
state $\a$ and $C_Q(K_{\a_1})$ is a cocycle factor; see {\it e.g.} 
Refs.~\cite{GHMR,GSW} for details). The dependence on the internal momenta 
looks different
in the two expressions (\ref{cartan1}) and (\ref{fgauge1}), but a moment
thought reveals that it is actually of the same form as required by
gauge invariance. Indeed, we can rewrite both equations 
in the following form
\begin{equation}
\Big( \cF_3^{\rm gauge} \Big)_{\a_1 \a_2}^{AB} = {g_{\rm{YM}} \over 
\sqrt{2}}\,
(\z_1 \cdot \z_2)\,A_{\mu}\, (k_1^{\mu}-k_2^{\mu}) \, \G^{AB}_{\a_1 \a_2}
\label{3ptgauge}
\end{equation}
where we have used the definitions of the normalization coefficients
to express the prefactor in terms of the Yang-Mills
coupling constant of the heterotic theory. Here
$\G^{AB}_{\a_1 \a_2}$ are the matrix elements of the antisymmetrized
product of two $\G$-matrices of $SO(32)$ which represent the fusion
coefficients among the adjoint 
and two spinor representations of $SO(32)$.~\footnote{See for instance
Ref.~\cite{KOST} for 
the expression of these
$\G$-matrices in terms of the internal momenta and cocycle factors.}

We are now in the position of evaluating the contribution to the 
diffusion amplitude among four particles
due to the exchange of gauge bosons at tree level.
In fact, this can be simply obtained by sewing two
3-point functions $\cF_3^{\rm gauge}$ with the massless propagator
(\ref{gaugebos}); in this way we obtain
\begin{equation}
\Big( \cF_4^{\rm gauge} \Big)_{\a_1 \a_4;\, \a_2 \a_3} = 
{g_{\rm{YM}}^2 \over 2}\,
(\z_1 \cdot \z_4)\,(\z_2 \cdot \z_3)\, {u-s \over t}\,
\left({1 \over 2}\,\G^{AB}_{\a_1\a_4}\G^{AB}_{\a_2\a_3}\right)
\label{fgauge4}
\end{equation}
where $s$, $t$ and $u$ are the usual Mandelstam variables which satisfy
$s+t+u = 16/\a'$. For later convenience, we introduce the 
adimensional variable $\sigma=-u\a'/4$, which in the limit $t\to 0$
is related to the Lorentz parameter $\g$ according to 
$\s=2(\g-1)$. Then, for $t\to 0$ Eq.~(\ref{fgauge4}) becomes
\begin{equation}
\Big( \cF_4^{\rm gauge} \Big)_{\a_1 \a_4;\, \a_2 \a_3}
=-\,g_{\rm{YM}}^2\,
(\z_1 \cdot \z_4)\,(\z_2 \cdot \z_3) \,{M^2_{\rm h}\,(2+\sigma)\over t}\,
\left({1 \over 2}\,\G^{AB}_{\a_1\a_4}\G^{AB}_{\a_2\a_3}\right)~~.
\label{fgauge3}
\end{equation}
Reverting to the standard field theory normalization by multiplying
each external leg by $1/\sqrt{2E}$, and removing for simplicity
the polarization factors, we finally obtain the gauge potential
energy
\begin{equation}
\Big(V_{\rm h}^{\rm gauge} \Big)_{\a_1 \a_4 ;\,\a_2 \a_3} 
= -\, {g_{\rm{YM}}^2 \over 2 }\,{1\over t}\,
\left({1 \over 2}\,\G^{AB}_{\a_1\a_4}\G^{AB}_{\a_2\a_3}\right)~~,
\label{fgauge41}
\end{equation}
which in configuration space becomes
\begin{equation}
\Big(V_{\rm h}^{\rm gauge}\Big)_{\a_1\a_4;\,\a_2\a_3}(r) 
= {g_{\rm YM}^2 \over 2} 
\,\left({1 \over 2}\,\G^{AB}_{\a_1\a_4}\G^{AB}_{\a_2\a_3}\right)\, 
{1 \over 7 \,\Omega_8 \,r^7}~~.
\label{gaugehet}
\end{equation}
Notice that this potential does not depend on the relative velocity
of the particles involved in the interaction; moreover, as expected,
it is a ``Coulomb-like'' potential for point particles in ten 
dimensions carrying the spinor representation of the gauge group.

We now turn to the gravitational interactions of the non-BPS
heterotic particles (\ref{alpha}) and (\ref{gamma}) following
the same steps we have described for the gauge interactions.
Let us recall that the massless bosonic states of the graviton multiplet
of the heterotic theory are 
\begin{equation}
\epsilon_{\m \n}\, \psi_{-1/2}^{\m}\, 
\tilde{\a}_{-1}^{\n} \ket{q} \otimes \ket{Q=0}\quad \mbox{with}~~q^2=0~~,
\label{massless}
\end{equation}
where the polarization is 
\begin{equation}
\epsilon_{\m \n} = h_{\m \n} = h_{\n \m} \quad , \quad
q^{\m}\, h_{\m \n}  = 0
\label{polgrav}
\end{equation}
for the graviton, 
\begin{equation}
\epsilon_{\m \n} = {\f \over \sqrt{8}}
\,\big(\eta_{\m \n} - q_{\m} \ell_{\n} - q_{\n} \ell_{\m} \big) \quad , \quad 
 q \cdot \ell = 1 \quad , \quad \ell^2 = 0
\label{poldil}
\end{equation}
for the dilaton, and 
\begin{equation}
\epsilon_{\m \n} = {1 \over \sqrt{2}}
\,B_{\m \n} =-{1 \over \sqrt{2}}
\,B_{\n \m}  \quad , \quad
q^{\m}\, B_{\m \n}  = 0
\label{polanti}
\end{equation}
for the antisymmetric Kalb-Ramond field.
The vertex operators corresponding to these states are written in Appendix C
in the $(-1)$ and $(0)$ superghost pictures, and will be denoted generically
by $\cV_{\rm grav}$. 

The gravitational coupling of the non-BPS particles can be determined
by evaluating the correlation function among two
vertex operators $\cV$ and one vertex operator $\cV_{\rm grav}$, namely
\begin{equation}
\Big( \cF_3^{\rm grav} \Big)_{\a_1\a_2} =
\cC_0\,  \hat{\cN}^3 \int d^2\m(z_i, \bar{z}_i) 
\,\Big\langle c{\bar c}\cV^{(-1)}_{1}(z_1,\bar{z}_1)\,{c\bar c} 
\cV^{(-1)}_{2}(z_2,\bar{z}_2)\,{ c\bar c}\cV^{(0)}_{\rm grav}(z_3,\bar{z}_3)
\Big\rangle ~~.
\label{ggrav}
\end{equation}
As before, also now we are interested only in the scalar part of this 
expression which is proportional to $\z_1\cdot\z_2$, since 
we want to compare it with the boundary state
calculation of \secn{boundarystate}. Using the expression
of the vertex operators reported in Appendix C, it is not difficult
to find that
\begin{equation}
\Big( \cF_3^{\rm grav} \Big)_{\a_1\a_2} = \cC_0 \,\hat{\cN}^3 \,\a'\,
\big (\z_1 \cdot \z_2 \big) \,\epsilon_{\m \n} \,k_2^{\m} \,k_2^{\n}~~,
\label{ggrav1}
\end{equation}
from which we read that the couplings of the non-BPS states
with the graviton, the dilaton and the
antisymmetric tensor field are 
\bea
\Big( \cF_3^{\rm grav} \Big)_{\a_1\a_2}^{(h)} &=& 4\kappa_{10}\,
(\z_1 \cdot \z_2)\,h_{\m\n} \,k_2^{\m} \,k_2^{\n}~~,
\nonumber \\
\Big( \cF_3^{\rm grav} \Big)_{\a_1\a_2}^{(\f)} &=& - 
\sqrt{2}\,\kappa_{10}\,M_{\rm h}^2\,(\z_1 \cdot \z_2) \,\phi
~~,
\label{dil1}\\
\Big( \cF_3^{\rm grav} \Big)_{\a_1\a_2}^{(B)} &=& 0~~.
\nonumber 
\eea
Note that the vanishing of the heterotic non BPS states coupling with the 
antisymmetric Kalb-Ramond field is consistent with the fact that the 
type I D-particle does not couple to the R-R 2-form.  

Now we can evaluate the diffusion amplitude of the non-BPS particles
due to gravitational exchanges by gluing two 3-point
functions (\ref{dil1}) with the appropriate massless propagators.
Summing over graviton and dilaton exchanges and 
introducing the same notation adopted for the gauge interactions,
we obtain
\begin{equation}
\Big({\cal F}_4^{\rm grav}\Big)_{\a_1\a_4;\,\a_2\a_3} 
= - (2\kappa_{10})^2\, 
(\z_1 \cdot \z_4)\,(\z_2 \cdot \z_3)\,{M_{\rm h}^4\,(2+\s)^2 \over t}~~.
\label{fgrav}
\end{equation}
Notice that, as expected, neither the 3-point functions (\ref{dil1}) nor the
4-point function (\ref{fgrav}) depend on the indices $\a_i$ that span 
the $SO(32)$ spinor representations carried by the non-BPS particles;
therefore they can be suppressed. Normalizing each external leg by a factor of 
$1/\sqrt{2E}$ and removing for simplicity the polarization terms, we can 
obtain from Eq. (\ref{fgrav}) the following gravitational potential energy 
\begin{equation}
V_{\rm h}^{\rm grav}(r) = (2 \kappa_{10})^2 \,
{M_{\rm h}^2 \,\g \over  7 \,\Omega_8\,r^7}~~,
\label{newtonhet}
\end{equation}
which in the small velocity limit becomes
\begin{equation}
V_{\rm h}^{\rm grav}(r) =  
(2 \kappa_{10})^2 \,{M_{\rm h}^2 \over 7\, \Omega_8\, r^7}\,
\left(1 + \shalf\, v^2\right) +o(v^2) ~~.
\label{newtonhet1}
\end{equation}
In Eq. (\ref{newtonhet}) we recognize Newton's law for point particles of mass
$M_{\rm h}$ separated by a distance $r$ in ten dimensions with the 
appropriate relativistic correction.

We conclude this section by mentioning that the same results 
(\ref{fgauge3}) and (\ref{fgrav}) can be directly obtained
by evaluating a 4-point function of non BPS states on the sphere, or more 
precisely its ``universal'' part in the $t$-channel which is proportional 
to $(\z_1\cdot\z_4)\,(\z_2\cdot\z_3)$.
In fact, using standard techniques, one can show that this part of the 4-point 
amplitude is  
\begin{equation}
\cA_4 = -\, {4 \over \pi^2}  \,\cC_0\,\hat{\cN}^4\, 
(\z_1 \cdot \z_4)\,( \z_2\cdot \z_3)\, 
 A(s,t,u;S,T,U)~~,
\label{ampl}
\end{equation}
where
\bea
A(s,t,u;S,T,U) &=& \!\eps(K)\,  \sin \Big[\pi \Big({s \a' \over 4} \Big) \Big]
\,\sin \Big[\pi \Big({t \a' \over 4} \Big) \Big]
\,\sin \Big[\pi \Big( {u \a' \over 4} \Big) \Big] \nonumber \\
&\times& \!\G \Big(3-{s\a' \over 4} \Big)\, \G \Big(-{t\a' \over 4} \Big)
\,\G \Big(3-{u\a' \over 4} \Big) \label{A}\\
&\times& \!\G \Big(-1-{s \a'  \over 4}-{S \over 2} \Big)\,
\G \Big(-1- {t \a' \over 4}-{T \over 2} \Big)\,
\G \Big(-1- {u \a' \over 4}-{U \over 2}\Big) ~.
\nonumber 
\eea
In this expression $S$, $T$ and $U$ are the Mandelstam variables for the
internal momenta which obey $S+T+U=-16$, and 
$\eps(K)\equiv c_{K_{3}}(K_1+K_2)\,c_{K_{2}}(K_1)(-)^{U/2}$ 
is a cocycle factor whose values are $\pm 1$ 
(see for instance Ref.~\cite{GSW}). Since we are interested only in the 
contributions due to exchanges of massless states in the $t$ channel, we must
look for the poles of $\cA_4$ with respect to $t$. Inspection of \eq{A} 
shows that these occur only for $T=0$ and $T=-2$. The poles 
for $T=0$ correspond to exchanges of gravitons, dilatons and
gauge bosons associated to the Cartan generators of $SO(32)$, while those 
for $T=-2$ correpond to exchanges of the remaining 480 gauge bosons.
In the limit $t\to 0$, we have
\begin{eqnarray}
\cA_4\,\Big|_{\,T=0} &\simeq& {16 \,\pi\, \cC_0\,\hat{\cN}^4  \over \a'} \,
\Big[(4+S/2)(2+\s) +(2+ \s)^2 \Big]\,{1\over t}~~,
\label{t=0}\\
\cA_4\,\Big|_{\,T=-2} &\simeq& -\,{16\, \pi\, \cC_0\,\hat{\cN}^4 \over \a' }\, 
(2+\s)\,{1\over t}~~.
\label{t=2}
\end{eqnarray}
We can disentangle the gauge and gravity pieces of \eq{t=0} by observing
that, because of gauge invariance, the gauge part at $T=0$ should have the 
same dynamical dependence as the amplitude (\ref{t=2}) for $T=-2$. 
Hence, we can conclude that the term of \eq{t=0} linear 
in $(2+\s)$ is due to gauge interactions, while the term quadratic
in $(2+\s)$ comes from gravity. Inspection of the coefficients
and a little algebra show that these expressions 
indeed match with Eqs. (\ref{fgauge3})
and (\ref{fgrav}), thus providing a strong check on our previous
calculations and on their interpretation. 
%\footnote{Observe that \eq{fgauge4} and \eq{fgrav}
%are consistent with respect to the factorization
%of closed-string amplitudes on the sphere since 
%$4\p \cC_0 \hat{\cN}^4 =  (\cC_0 \hat{\cN}^3)^2 \a'$}.

\vskip 1.5cm
\sect{Conclusions}
\label{conclusions}
\vskip 0.5cm

We now compare the results obtained in the previous
sections and discuss their relation in the light of the heterotic/type I
duality. For the gravitational interactions, 
the comparison is quite simple since in both theories we have found a 
potential energy of the form
\begin{equation}
V^{\rm grav}(r) =  (2 \kappa_{10})^2\,{M^2 \over 7\, \Omega_8\, r^7}\,
\left(1 + \shalf\, v^2\right) +o(v^2)
\label{newtonT}
\end{equation}
in the non-relativistic limit (see Eqs. (\ref{newton1}) and 
(\ref{newtonhet1})).
The only thing that one has to do to have complete agreement
is to change the values of the 
gravitational coupling constant $\kappa_{10}$ and of 
the mass $M$ according to the duality map as we discussed in the introduction.
What is nice to observe is that these changes make the two
gravitational potential energies agree not only at the static level
but also at the first non-trivial order in the velocity $v$. 

For the gauge interactions the situation is a bit different. Both
in the type I theory and in the heterotic string we have found that the
gauge potential energy of the stable non-BPS states is in the
form of Coulomb's  law (see Eqs. (\ref{vgauge0}) and (\ref{gaugehet})).
However, the detailed gauge group structure is not the
same in the two cases. The reason for this is quite simple. In the
heterotic theory one is able to describe the non-BPS particles in a complete
way because they are perturbative configurations of the heterotic string, and
in particular one can fully specify the polarizations of these
states also with respect to the gauge group. This is why the gauge
amplitudes involving these particles explicitly depend on the indices of the 
spinorial representation of $SO(32)$ (see Eqs. (\ref{3ptgauge}) and 
(\ref{fgauge3})). On the other hand, in the type I theory the 
non-BPS particles are non-perturbative configurations of the type I
string, and thus the description one is able to provide for them using
perturbative methods is necessarily incomplete.
This fact should not be surprising, because also in the case of the 
supersymmetric BPS D-branes one is not able to account for their
degeneracy (with respect to both the Lorentz group and the gauge group) 
using open strings with Dirichlet boundary conditions or equivalently boundary
states. Indeed with these methods one can compute
only the ``universal'' parts of the interactions involving
D-branes.

In Section 2 we have introduced a method to describe the emission of a 
colored gauge boson from a non-BPS D-particle viewed as a source 
carrying not the spinorial indices of $SO(32)$, but rather 
those of the bi-fundamental representation
formed with the Chan-Paton factors of the boundary changing vertex
operators $\cV_{09}$ and $\cV_{90}$ (see Eqs. (\ref{r09}) and 
(\ref{r90})).
In this framework, using the various kinds of open strings of type I 
we have been able to 
account for the gauge interactions of the non-BPS D-particles, but then the
comparison with the heterotic theory is not immediate.
In order to do such a comparison, we must ``reduce'' the heterotic
gauge potential energy by taking into account the contribution 
of all pairs of states compatible with the emission of a gauge boson of 
definite color. {F}rom the group theory point of view, this
amounts to transform the spinorial indices of
$\Big(V_{\rm h}^{\rm gauge}\Big)_{\a_1\a_4;\,\a_2\a_3}$ given in
Eq. (\ref{gaugehet})
into those of the bi-fundamental representation.
This can be easily done by noting that
\begin{equation}
\Big( \G^A \,\G^D\Big)^{\a\b}\,\Big(\G^{BC}\Big)_{\b\a}  
= \rm{Tr}\Big( \G^A \,\G^D\,\G^{BC} \Big) 
= \rm{Tr}\Big(\!\!\one\Big)\,\Big(\d^{BD}\,\d^{AC}-\d^{CD}\,\d^{AB}\Big)~~.
\end{equation}
Then, using this identity and Eq. (\ref{gaugehet}), we obtain the following 
reduced gauge potential energy for the heterotic non-BPS particles
\begin{eqnarray}
V_{\rm h}^{\rm gauge}(r) &\equiv& {\rm{Tr}}\Big(\!\!\one\Big)^{-2}
\Big( \G^A\, \G^D\Big)^{\a_4\a_1} \Big( \G^E\, \G^F\Big)^{\a_3\a_2}\,
\Big(V_{\rm h}^{\rm gauge}\Big)_{\a_1\a_4;\,\a_2\a_3}(r)
\nonumber \\ 
&=&
-\, \frac{g_{\rm YM}^2}{2}~\left(\delta^{AE}\,\delta^{FD}
-\delta^{AF}\,\delta^{DE}\right)~\frac{1}{7\,\Omega_8\,r^7}~~.
\end{eqnarray}
This expression exactly agrees with the corresponding one for the type I
theory given in Eq. (\ref{vgauge0}).

We remark that it is not meaningful to perform this reduction
directly on the 
heterotic 3-point function (\ref{3ptgauge}) and then compare it 
with the type I 
amplitude (\ref{f32}) describing the emission of a gauge boson from a
D-particle. In fact, in the perturbative type I theory the D-particle 
is an infinitely massive object which acts as a reservoir of 
momentum, and thus the space-time structure of its amplitudes cannot match 
with that of the heterotic scattering amplitudes. 
In other words, the diagram represented in Figure 1 
describing the gauge emission
from a D-particle of type I must not be considered as a vertex 
or a 3-point function in the field theory sense, but rather as a 1-point 
function in some definite background. A similar situation occurs
also in the gravitational sector where the boundary state representing the
D-particle generates all its 1-point functions, {\it i.e.} all its couplings 
with the closed string states of the bulk. In this sense, what we have done 
in Section 2 is to find the 1-point function of the 
non-BPS D-particle with the massless states of the other sector of the bulk, 
namely the open strings of type I. It would be nice to extend these results 
to all states of this open string sector. 
 
In conclusion, in this paper we have described how to compute the gauge and 
gravitational potential energies of the non-BPS D-particles of type I and 
shown that these agree with the corresponding ones computed for
the dual heterotic states provided that one uses the known duality and
renormalization effects. Our results thus provide a dynamical test
of the heterotic/type I duality at the non-BPS level.

\vskip 2cm
{\large {\bf {Acknowledgments}}}
\vskip 0.5cm
We would like to thank C. Bachas, M. Bill\`o, P. Di Vecchia, M. Frau, B. 
Pioline, C. Schweigert and R. Russo for several useful discussions. We 
especially thank B. Pioline for valuable discussions and remarks.

\vskip 2cm
\noindent
\appendix{\Large{\bf {Appendix A}}}
\renewcommand{\theequation}{A.\arabic{equation}}
\setcounter{equation}{0}
\vskip 0.5cm

\noindent
In this appendix we describe the
gauge emission from a D-string of type I,
following the scheme we proposed in Section 2.

As a consequence of the BPS condition, two parallel BPS D$p$-branes
of type I or type II do not exert any force on each other.
In the type II theories, the interaction between two branes is mediated only 
by the exchange of closed strings and the vanishing of the force
is easily seen using boundary states. Indeed one finds that
\begin{equation}
\bra{Dp} P \ket{Dp} =0
\label{a1}
\end{equation}
at the leading order in the string coupling constant.
In the limit of large distance between the branes, 
when only massless closed string states are exchanged, this means 
that the attraction due to gravitons and dilatons is compensated by the 
repulsion due to the $p+1$ R-R form under which the D$p$-branes are charged. 
In the case of two parallel D-strings
the interaction (\ref{a1}) is globally invariant under the world-sheet parity
$\Omega$ so that it vanishes also in type I theory.  However, in this 
theory one has to consider also the exchange of open 
9-9 strings which are present in the bulk and whose first contribution -- 
associated to a disk with two boundary components on the D-strings 
and two on the 
D9 branes -- appears at the next-to-leading 
order in the string coupling constant. 
For the no force condition to be true, this disk amplitude 
has thus to vanish identically. 
At first sight, this seems striking since the D-string is charged under 
the gauge potential (in fact it carries the $SO(32)$ spinorial 
representation). However, one must recall that, being an extended object, 
the D-string cannot be minimally coupled to the gauge potential and thus 
the naive conclusion does not apply. 
The aim of this appendix is to 
evaluate the gauge emission from a D-string of type I
using the same methods applied in the case of the D-particle 
(but without the technicalities due to the boundary fermion $\eta$),
and then to compute its contribution to the 
diffusion process between two D-strings.

We first briefly discuss the spectrum of the 1-9 open strings stretching 
between a D-string and a D9 brane. As for the 0-9 strings, also here
the NS sector is massive and does not represent any
degeneracy of the D-string; thus we do not consider it. In the R sector, 
instead, there are massless states. Since the world-sheet fermions 
$\psi^0$, $\psi^1$ have zero modes, the massless R ground state 
is a GSO even (chiral) spinor of $SO(1,1)$. The corresponding vertex 
operator reads
\begin{equation}
\cV_{91}^{(-1/2)} = \l^A\,S^+\, \D'\, \ee^{- \f/2}
\end{equation}
where $S^+$ is a positive chirality spin field of conformal dimension 1/8, and
$\D'$ is a boundary changing operator for the eight space directions
transverse to the D-string of conformal dimension $1/2$.
The vertex operator for the massless R states of the 1-9 sector is 
obtained by acting with $\Omega$ on $\cV_{91}$; thus
\begin{equation}
\cV_{19}^{(-1/2)}=\Omega\,\cV_{91}^{(-1/2)} = {\l^t}^{\,A}\,S^+ 
\,\D'\, \ee^{- \f/2}~~.
\end{equation}

We now evaluate the coupling of the D-string with the gauge bosons of type I.
As in the case of the D-particle, this is determined by the amplitude
on a disk with one boundary component on the D-string and one component
on one of the 32 D9-branes 
from which a gauge boson is emitted, and is represented by the 1-point
function of the gauge boson in the vacuum representing the D-string, {\it i.e.}
\begin{equation}
{\cal F}^{\rm gauge} = 
\langle \Omega_{19} \vert V^{(-1)}_{\rm 
gauge}(1) \vert \Omega_{91} \rangle
\end{equation}
where 
\begin{equation}
\vert \Omega_{19} \rangle 
= \lim_{z\rightarrow 0} \cV^{(-1/2)}_{19}(z) \vert 0 \rangle
~~~{\rm and}~~~\langle \Omega_{19}\vert =
\lim_{z\to \infty}\bra{0}\,\cV^{(-1/2)}_{19}(z)~~.
\end{equation}
After introducing the appropriate normalization factors, we can express
$\cF^{\rm gauge}$ as 
\begin{equation}
{\cal F}^{\rm gauge} = 
{\cal C}_{\rm disk}~{\cal N}^2_\R ~{\cal N}_\NS~
\int d\mu(z_i) ~\left
\langle 
c\,\cV^{(-1/2)}_{19}(z_1)~c\,\cV^{(-1)}_{\rm 
gauge}(z_2)~c\,\cV^{(-1/2)}_{91}(z_3) \right \rangle ~~.
\label{a4}
\end{equation}
For the same reasons discussed in the case of the D-particle, also here
there is no emission of gauge fields with polarizations along the directions 
transverse to the D-string; thus we have emissions only in the two
longitudinal directions $\mu=0,1$, but these can occur with arbitrary
transverse momentum.
To evaluate (\ref{a4}) we need the following basic correlators \cite{KOST}
\begin{eqnarray}
\langle \Delta'(z_1)\, \ee^{\ii k\cdot X(z_2)}\,\Delta' (z_3)\rangle
&=& z_{12}^{-\a' k^2}\,z_{13}^{\a' k^2-1}\,z_{23}^{-\a' k^2}~~,
\nonumber\\
\langle S^{+}(z_1)\, \psi^{\mu}(z_2) \,S^{+}(z_3)\rangle &=& 
(\Gamma^{\mu}C^{-1})^{++}\,z_{12}^{-{1/ 2}}\,z_{13}^{{1 /4}}
\,z_{23}^{-{1 / 2}}
\end{eqnarray}
for $\mu=0,1$.
Inserting them into Eq. (\ref{a4}), we obtain
\begin{eqnarray}
{\cal F}^{\rm gauge} &=& {\cal C}_{\rm disk}~{\cal N}^2_\R ~{\cal N}_\NS~
{\rm Tr}
({\l^t}^{\,A}\,\Lambda^{BC}\,\lambda^D)
~ A_{\mu}\,(\Gamma^{\mu}C^{-1})^{++}
\,\int d\mu(z_i)
\,\left({z_{13}\over z_{12}\,z_{23}} \right)^{\a' k^2} 
\nonumber \\ 
&=& \ii\,g_{\rm YM}\,(\delta^{AB}\delta^{CD}-\delta^{AC}\delta^{BD})
\,(A_{0}- A_1)
\end{eqnarray} 
where we have used that $(\Gamma^{0}C^{-1})^{++} = 
-(\Gamma^{1}C^{-1})^{++} =1$. As anticipated, this is not a minimal gauge 
coupling because the D-string is an extended object.

Now, using this coupling and the propagator (\ref{gaugebos}) for the massless 
gauge bosons, we can obtain the gauge potential energy 
$V_{\rm I}^{\rm gauge}$ between two D-strings given by
$V_{\rm I}^{\rm gauge} \approx \cF^{\rm gauge} \cP \cF^{\rm gauge}$.
Inserting the explicit values, we find 
\bea
V_{\rm I}^{\rm gauge} = g_{\rm YM}^2\,\Big(\delta_{AE}\,\delta_{DF}-
\delta_{AF}\,\delta_{DE}\Big) 
\,\frac{(-1+1)}{q^2} = 0~~.
\label{agaugeI}
\eea
The vanishing of this contribution confirms our previous statements
about the no force condition.

\vskip 1.5cm
\appendix{\Large{\bf {Appendix B}}}
\renewcommand{\theequation}{B.\arabic{equation}}
\setcounter{equation}{0}
\vskip 0.5cm

\noindent
In this appendix we present the definitions of the
various normalization factors that are needed for the calculations
presented in Sections 2, 3 and 4.

The gravitational coupling
constant $\kappa_{10}$, the normalization $\cC_{0}$
of the closed-string tree-level diagrams (sphere diagrams) and
the normalization factor $\hat{\cN}$ of the closed string vertex
operators have the same expressions for both the $SO(32)$ heterotic
string and the type I theory, and are given by
\bea
\kappa_{10} &=& 8\, \p^{7/2}\, \a'^{\,2}\, g ~~,
\label{k10} \\
\cC_{\rm 0} &=& (2\p)^{-4}\, \a'^{\,-5}\, g^{-2}~~,
\label{C0} \\
\hat{\cN} &=& 8\, \p^{5/2}\, \a'^{\,2} \,g~~,
\label{nclosed}
\eea
where $g$ is the string coupling constant ($g_{\rm h}$ for the heterotic
string and $g_{\rm I}$ for the type I theory).

In the heterotic string, the gauge coupling constant $g_{\rm YM}$
is related to $\kappa_{10}$ as follows \cite{pol}
\begin{equation}
 g_{\rm YM}^2={4 \over \a'}~ \kappa_{10}^2 
= 2^8\,\pi^7\,\a'^{\,3}\,g_{\rm h}^2~~,
\label{YMhet}
\end{equation}
while in the type I theory the relation between $g_{\rm YM}$
and $\kappa_{10}$ is \cite{pol}
\begin{equation}
 g_{\rm YM}^2={2^{3/2} \over \a'\,g_{\rm I}}~ \kappa_{10}^2 
= 2^{15/2}\,\pi^7\,\a'^{\,3}\,g_{\rm I}~~.
\label{YMI}
\end{equation}
In type I string theory one must consider also diagrams involving
open strings. The normalization of the disk diagrams is \cite{pol,1loop}
\begin{equation}
\cC_{\rm disk} = g_{\rm YM}^{-2} \,(2 \a')^{-2}~~,
\label{Cdisk}
\end{equation}
while the normalization factors of the open string vertex operators
in the NS and R sectors are
\bea
\cN_{\NS} &=& g_{\rm YM}\, \sqrt{2 \a'} \nonumber \\
\cN_{\R} &=& g_{\rm YM} \,(2 \a')^{3/4}~~.
\label{nopen}
\eea

We now list the expression of the various
Chan-Paton factors that were used in
Section 2. The factor $\Lambda^{AB}$
carried by the gauge boson
has indices in the adjoint representation
of $SO(32)$.
Its matrix elements explicitly read
\begin{equation}
(\L^{AB})_{CD} = {\ii} \,\Big(\d_C^A\,\d_D^B - \d_D^A\,\d_C^B\Big)~~.
\label{Lambda}
\end{equation}
The Chan-Paton factor $\lambda^A$ associated to 0-9 strings is a column vector 
with an index in the fundamental representation of $SO(32)$.
It reads
\begin{equation}
(\l^A)_B = \d_B^A~~.
\label{lamdba}
\end{equation}
With these expressions it is easy to see that
\bea
{\rm Tr}\Big( \L^{AB}\, \L^{CD} \Big) &=& 2\,\Big(\d^{AC}\, \d^{BD} - 
\d^{AD}\, \d^{BC}\Big)~~, \nonumber  \\
{\rm Tr}\Big({\l^t}^{\,A} \L^{BC} \l^D\Big) &=& \ii\, \Big(\d^{AB}\, \d^{CD} 
- \d^{AC}\,\d^{BD}\Big)~~.
\eea

\vskip 1.5cm 
\appendix{\Large{\bf {Appendix C}}}
\renewcommand{\theequation}{C.\arabic{equation}}
\setcounter{equation}{0}
\vskip 0.5cm
\noindent
In this appendix we write 
the vertex operators of the non-BPS heterotic
states (\ref{alpha}) and (\ref{gamma}), of the
gauge bosons of $SO(32)$ given in Eqs. (\ref{cartan}) and (\ref{charged}),
and of the bosonic states (\ref{massless}) of the graviton multiplet.

In the $(-1)$ superghost picture, the vertices associated to
the non-BPS states (\ref{alpha}) and (\ref{gamma}) are
\bea
\cV^{(-1)}_\cA(z, \bar{z}) &=& \z_{\m}\, 
\cA^{\m}(z)\,
\ee^{\ii k \cdot X(z , \bar{z})}\,\, C_K\,
\ee^{\ii {K \over \sqrt{2\a'}} \cdot \tilde{X}(\bar{z})} ~~,
\label{vertexa1} \\
\cV^{(-1)}_\cB(z, \bar{z}) &=& - \,{1 \over \sqrt{2 \a'}}\, \z_{\m \n}\,
\cB^{\m \n}(z) \,\ee^{\ii k  \cdot X(z , \bar{z})} \,\,
C_K\,
\ee^{\ii {K \over \sqrt{2\a'}} \cdot \tilde{X}(\bar{z})}~~, 
\label{vertexb1} \\
\cV^{(-1)}_\cC(z, \bar{z}) &=& {1 \over 3!} \,
\z_{\m \n \r}\,\cC^{\m \n \r}(z)\, \ee^{\ii k  \cdot X(z, \bar{z})}
\,\,C_K\,
\ee^{\ii {K \over \sqrt{2\a'}} \cdot \tilde{X}(\bar{z})} ~~,
\label{vertexc1}
\eea
where 
\bea
\cA^{\m}(z) &=&  \partial \psi^{\m} (z)\,\ee^{-\f(z)}~~,
\nonumber \\
\cB^{\m \n}(z) &=&   \psi^{\m} (z)\,\partial x^{\n}(z)
\,\ee^{-\f(z)}~~, \label{-1} \\ 
\cC^{\m \n \r}(z) &=&  \psi^{\m} (z)\, 
 \psi^{\n} (z) \,\psi^{\r} (z)\,\ee^{-\f(z)}~~.
\nonumber
\eea
In Eq. (\ref{vertexb1}) the polarization tensor $\z^{\m \n}$
is symmetric or antisymmetric depending on whether one considers
a state in the {\bf 44} or in the {\bf 84} representation of the Lorentz 
group. Moreover, in all vertex operators we have introduced 
suitable cocycle factors 
$C_K$, which depend only on the internal momenta and 
satisfy \cite{GSW}
\begin{equation}
C_K(P)\, C_{K'}(P) = C_{K+K'}(P)~~.
\label{cocycle}
\end{equation}
Applying the picture-changing operator to $\cV^{(-1)}$
we can obtain the vertices $\cV^{(0)}$ in the
$(0)$ superghost picture. They are given by the same expressions
(\ref{vertexa1}) -- (\ref{vertexc1}) with $\cA^{\m}$, $\cB^{\m \n}$
and $\cC^{\m \n \r}$ replaced respectively by
\bea
\hat\cA^{\m}(z) &=& \! {\ii \over \sqrt{2 \a'}} 
\left[ \partial^2X^{\m}(z) - \ii \a' 
  (k \cdot \psi ) \partial \psi^{\m}(z) \right]~~, \nonumber  \\
\hat\cB^{\m \n}(z) &=& \!\ii \sqrt{2 \a'} \Big[ \psi^{\m} (z)
\partial \psi^{\n}(z) - (\ii/2) (k \cdot \psi) 
\psi^{\m} (z)\partial X^{\n}(z) + {1 \over 2 \a'} 
\partial X^{\m}(z)\partial X^{\n}(z) \Big] \nonumber~, \\
\hat\cC^{\m \n \r}(z) &=&\! {\ii \over \sqrt{2 \a'}} \Big[- \ii \a' (k 
\cdot \psi) \psi^{\m} (z)\psi^{\n} (z)\psi^{\r} (z) +  
 \partial X^{\m}(z)
\psi^{\n} (z)\psi^{\r}(z)   \nonumber \\
  &&-\,\psi^{\m}(z) \partial X^{\n}(z)
 \psi^{\r}(z) +\psi^{\m} (z)\psi^{\n}(z)\partial X^{\m}(z) \Big]~~.
\label{0}
\eea
The vertex operators for the gauge bosons (\ref{cartan})
associated to the 16 Cartan generators of $SO(32)$ in the $(-1)$ superghost
picture are  
\begin{equation}
\cV_{\rm gauge}^{(-1)}(z,\bar{z}) = {\ii  \over \sqrt{2 \a'}} 
\, A_{\m} \,\psi^{\m}(z)\,\ee^{- \f(z)}\,  \ee^{\ii q \cdot X(z, \bar{z})} 
 \,\,\bar{\partial} \tilde{X}^I(\bar{z})~~,
\end{equation}
while those for the gauge bosons (\ref{charged}) associated to the
480 remaining generators are
\begin{equation}
\cV_{\rm gauge}^{(-1)}(z,\bar{z}) =   A_{\m}\,
\psi^{\m}(z)\,\ee^{- \f(z)}\, \ee^{\ii q \cdot X(z, \bar{z})}\,\, C_Q 
\ee^{\ii {Q \over \sqrt{2\a'}} \cdot \tilde{X}(\bar{z})}~~.
\label{boson-1}
\end{equation}
In the $(0)$ superghost picture these vertices become respectively
\begin{equation}
\cV_{\rm gauge}^{(0)}(z,\bar{z}) = -\,{1 \over 2 \a'}\,
 \,A_{\m} \Big[\partial X^{\m}(z)  - \ii\,\a'
\, (q \cdot \psi) \,\psi^{\m}(z) \Big] \,\ee^{\ii q \cdot X(z, \bar{z})} 
\,\, \bar{\partial} \tilde{X}^I(\bar{z}) ~~,
\end{equation}
and 
\begin{equation}
\cV_{\rm gauge}^{(0)}(z,\bar{z}) = {\ii  \over \sqrt{2 \a'}}\,
  A_{\m}\,
\Big[\partial X^{\m}(z)  - \ii\,\a'
 \,(q \cdot \psi) \,\psi^{\m}(z) \Big] \,\ee^{\ii q \cdot X(z, \bar{z})} 
\,\, C_Q \,\ee^{\ii {Q \over \sqrt{2\a'}} \cdot \tilde{X}(\bar{z})}~~.
\label{boson0}
\end{equation}
Finally, the vertices for the bosonic states (\ref{massless}) 
of the graviton multiplet are
\begin{equation}
\cV_{\rm grav}^{(-1)}(z,\bar{z}) = {\ii  \over \sqrt{2 \a'}}\,
\epsilon_{\m \n} \, \psi^{\m}(z)\, \ee^{- \f(z)}\,
\bar{\partial} \tilde{X}^{\n}\,
(\bar{z}) \ee^{\ii q \cdot X(z, \bar{z})}
\label{vertexgrav}
\end{equation}
in the $(-1)$ superghost picture, and 
\begin{equation}
\cV_{\rm grav}^{(0)}(z,\bar{z}) = -\,{1 \over 2 \a'}\,
\epsilon_{\m \n}\, \Big[ \partial X^{\m}(z) - \ii \,\a'\, (q \cdot \psi)\,
\psi^{\m}(z) \Big]\,\bar{\partial} \tilde{X}^{\n}
(\bar{z})\, \ee^{\ii q \cdot X(z, \bar{z})}
\label{vertexgrav0}
\end{equation}
in the $(0)$ superghost picture.

%%%%%%%

\end{document}